\documentclass[showpacs]{revtex4}

\begin{document}

\preprint{arXiv:1106.0610v2 [hep-ph]}

\title{Comment on : A Proposal to Measure Photon-Photon Scattering}

\author{Denis Bernard}
\affiliation{Laboratoire Leprince-Ringuet, Ecole Polytechnique, CNRS/IN2P3, F-91128 Palaiseau, France }
 
\date{\today}%

\begin{abstract}
I comment on a recent preprint ``A Proposal to Measure Photon-Photon
Scattering'' that appeared recently as arXiv:1106.0465v1 [hep-ph].
\end{abstract}


\maketitle

The Authors of Ref. \cite{Fujita:2011rd} proposed that the low-energy
expression for the cross section of photon-elastic scattering
(Heisenberg and Euler \cite{euler}) : 

\begin{equation}
 {d\sigma\over d\Omega}\simeq {139\alpha^4\over (180\pi)^2 m^2} 
\left({\omega\over m} \right)^6 (3+\cos^2\theta )^2 
\end{equation}

, be replaced by (\cite{kanda}, to be published) : 

\begin{equation}
 {d\sigma\over d\Omega} \simeq {\alpha^4\over 
(12\pi)^2 \omega^2} (3+2\cos^2\theta +\cos^4\theta ) 
 \label{eq:kanda}
\end{equation}

where $\omega$ and $m$ denote the photon energy and the mass of
electron, respectively.
The orders of magnitude that were given  for $\omega \simeq 1$ eV are 
$2.3 \times 10^{-21}  \ \ {\rm cm}^2 $
and 
$9.3 \times 10^{-67} \ \ {\rm cm}^2 $, 
respectively.

~

Experimentally, an upper limit was  obtained first by colliding two
laser beams head-on, and searching for one of the scattered photons
(Moulin {\em et al.}  \cite{Moulin:1996vv}) :
\begin{equation}
\sigma <  10^{-39} \ \ {\rm cm}^2.  
\label{eq:direct}
\end{equation}

The limit was further improved with a stimulation of the reaction with
a third beam (Bernard {\em et al.} \cite{Bernard:2010dx}), based on
the computations of Moulin {\em et al.}  \cite{Moulin:2002ya} 
(that were revisited recently by Lundstrom, Lundin {\em et al.}
\cite{Lundstrom:2005za,Lundin:2006wu}).

\begin{equation}
\sigma < 1.5 \times 10^{-48} \ \ {\rm cm}^2. 
\label{eq:stimulated}
\end{equation}

These two experimental publications
\cite{Moulin:1996vv,Bernard:2010dx} are indeed mentionned in
Ref. \cite{Fujita:2011rd}, but it would be interesting to understand
how the prediction of eq. (\ref{eq:kanda}) can face the upper limits
of eqs. (\ref{eq:direct}) and (\ref{eq:stimulated}).

~

Also, it should be noted that the QED computation of
the 4-photon loop diagrams  (See, e.g.,
\cite{Kinoshita:2002ns} and a review at \cite{Jegerlehner:2009ry}) take part into the prediction of the lepton
``anomalous''  magnetic moment, which is found to be in fair (i.e., within three
standard deviations) agreement with the experimental measurement (e.g.
\cite{Davier:2010nc}).

\end{document}